\documentclass{amsart}
\usepackage{epsfig}
\usepackage{amsmath}
\usepackage{amsfonts}
\usepackage{amssymb}
\usepackage{euscript}
\usepackage{graphicx}
\begin{document}

\sloppy

\sloppy
\begin{center}
{{
%\Huge%\bfseries
Superfluidity and zero-point oscillations
}}
\vspace{1cm}

Boris Vasiliev,\\

Russia\\

{E-mail address: bv.vasiliev@narod.ru}\\
\end{center}

%-------------------------------------------------------------------------------
%------------------------

%\tableofcontents

\section{The Introduction}
%{Superfluidity as a sequence of  ordering of zero-point oscillations}

It has been shown  recently that the origin  of superconductivity can be in an ordering zero-point oscillations \cite{BV1}-\cite{BV3}.
Superfluidity is a related phenomenon.
So it makes sense to consider the role of ordering of zero-point oscillations in the emergence of superfluidity.

%\subsection{Zero-point oscillation of He atoms and superfluidity}

The main features of superfluidity of liquid helium became clear few decades ago
\cite{Landau}, \cite{Halat}. L.D.Landau explains this phenomenon as the manifestation of a quantum behavior of the macroscopic object.

However, the causes and mechanism of the formation of superfluidity  are not clear till our days.
There is no explanation why the $\lambda$-transition in helium-4 occurs at about 2 K, that is  about twice less than   its boiling point:
\begin{equation}
\frac{T_{boiling}}{T_\lambda}\approx 1.94,\label{f022}
\end{equation}
while for helium-3, this transition is observed only at temperatures  about a thousand times smaller.

 The related phenomenon, superconductivity, can be regarded as superfluidity of a charged liquid. It can be quantitatively described considering it as the consequence of ordering of zero-point oscillations of electron gas. Therefore it seems appropriate to consider superfluidity from the same point of view.

The atoms in liquid helium-4 are electrically neutral, as they have no dipole moments and do not form molecules.
  Yet some electromagnetic mechanism should be responsible for phase transformations of liquid helium (as well as in other condensed substance where phase transformations are related to the changes of energy of the same scale).

 F. London has demonstrated already  in the 1930's \cite{FLondon}, that there is an interaction between the atoms in the ground state, which has a quantum nature. It can be considered as a kind of  the Van-der-Waals interaction.
 Atomic shells in s-state perform zero-point oscillations. F.London was considering  vibrating atomic shells as three-dimensional oscillating dipoles which  are connected to each other by the electromagnetic interaction. He called this interaction of atoms in the ground state the dispersion interaction.

\section{The dispersion effect in interaction of atoms in the ground state}
Following F.London \cite{FLondon}, let us consider two spherically symmetric atoms without non-zero average dipole moments.  Let us suppose that at some time the charges of these atoms are  fluctuationally displaced from the equilibrium states:
\begin{equation}
r_1=(x_1,y_1,z_1)\nonumber
\end{equation}
\begin{equation}
r_{2}=(x_{2},y_{2},z_{2})\nonumber
\end{equation}
If atoms are located along the Z-axis at the distance $L$ of each other, their potential energy can be written as:
\begin{equation}
\mathcal{H}=\underbrace{\frac{e^2 r_1^2}{2A}+\frac{e^2 r_{2}^2}{2A}}_{{elastic~
dipoles~energy}}
+\underbrace{\frac{e^2}{L^3}(x_1 x_{2}+y_1y_{2}-2z_1z_{2})}_{elastic~dipoles~interaction},\label{q0}
\end{equation}
where $A$ is the polarizability of atom.

The Hamiltonian can be diagonalized by using the normal coordinates of symmetric and antisymmetric displacements:
\begin{displaymath}
r_s\equiv
\left\{
\begin{array}{ll}
x_s=\frac{1}{\sqrt{2}}(x_1+x_2) \\
y_s=\frac{1}{\sqrt{2}}(y_1+y_2) \\
z_s=\frac{1}{\sqrt{2}}(z_1+z_2) \\
\end{array}
\right.
\end{displaymath}
and
\begin{displaymath}
r_a\equiv
\left\{
\begin{array}{ll}
x_a=\frac{1}{\sqrt{2}}(x_1-x_2) \\
y_a=\frac{1}{\sqrt{2}}(y_1-y_2) \\
z_a=\frac{1}{\sqrt{2}}(z_1-z_2) \\
\end{array}
\right.
\end{displaymath}
This yields
\begin{displaymath}
\begin{array}{ll}
x_1=\frac{1}{\sqrt{2}}(x_s+x_a) \\
y_1=\frac{1}{\sqrt{2}}(y_s+y_a) \\
z_1=\frac{1}{\sqrt{2}}(z_s+z_a) \\
\end{array}
\end{displaymath}
and
\begin{displaymath}
\begin{array}{ll}
x_2=\frac{1}{\sqrt{2}}(x_s-x_a) \\
y_2=\frac{1}{\sqrt{2}}(y_s-y_a) \\
z_2=\frac{1}{\sqrt{2}}(z_s-z_a) \\
\end{array}
\end{displaymath}
As the result of this change of variables we obtain:
{\scriptsize
\begin{eqnarray}
\mathcal{H}=\frac{e^2}{2A}(r_s^2+r_a^2)+
\frac{e^2}{2L^3}(x_s^2+y_s^3-2z_s^2-x_a^2-y_a^2+2z_a^2)= \nonumber\\
=\frac{e^2}{2A}\left[\left(1+\frac{A}{L^3}\right)(x_s^2+y_s^2)+ % \nonumber
\left(1-\frac{A}{L^3}\right)(x_a^2+y_a^2)+ \right. \\ %\nonumber
+\left(1-2\frac{A}{L^3}\right)z_s^2 +
\left.\left(1+2\frac{A}{L^3}\right)z_a^2\right].\nonumber\label{q01}
\end{eqnarray}}
Consequently, frequencies of oscillators depend on their  orientation and they are determined by the equations:
{\scriptsize
\begin{eqnarray}
\Omega_{0x}^{s\atop a}=\Omega_{0y}^{s\atop a}=\Omega_0\sqrt{1\pm\frac{A}{L^3}}\approx\Omega_0\left({1\pm\frac{A}{L^3}}-\frac{A^2}{8L^6}\pm...\right)  \\
\Omega_{0z}^{s\atop a}=\Omega_0\sqrt{1\mp\frac{2A}{L^3}}\approx\Omega_0\left({1\mp\frac{A}{L^3}}-\frac{A^2}{2L^6}\mp...\right),
\end{eqnarray}}
where
\begin{equation}
\Omega_0=\frac{2\pi e}{\sqrt{mA}}%\nonumber
\end{equation}
is natural frequency of the electronic shell of the atom (at $L\rightarrow\infty$).
The energy of zero-point oscillations is
\begin{equation}
\mathcal{E}=\frac{1}{2}\hbar(\Omega_0^s+\Omega_0^a).%\label{EE}
\end{equation}
It is easy to see that the description of interactions between neutral atoms do not contain terms $\frac{1}{L^3}$, which are characteristics for the interaction of zero-point oscillations in the electron gas  and
which are responsible for the occurrence of superconductivity \cite{BV3}. \\
The terms that are proportional to $\frac{1}{L^6}$ manifest themselves in interactions of neutral atoms.\\

It is important to emphasize that the energies of interaction are different for  different orientations of zero-point oscillations. So the interaction of zero-point oscillations oriented along the direction connecting the atoms leads to their attraction with energy:
\begin{equation}
\mathcal{E}_z = - \frac{1}{2}\hbar\Omega_0\frac{A^2}{L^6},\label{f1}
\end{equation}
while the summary energy of the attraction of the oscillators of the perpendicular directions (x and y)  is equal to one half of it:
\begin{equation}
\mathcal{E}_{x+y} = - \frac{1}{4}\hbar\Omega_0\frac{A^2}{L^6}\label{f2}
\end{equation}
(the minus sign is taken here because  for this case   the opposite direction of dipoles  is energetically favorable).

\section{The estimation of  main characteristic parameters of superfluid helium}
\subsection{The main characteristic parameters of the zero-point oscillations of atoms in superfluid helium-4}
There is no repulsion in a gas of neutral bosons.
Therefore, due to attraction between the atoms at temperatures below
\begin{equation}
T_{boil}=\frac{2}{3k}\mathcal{E}_z%\label{f1}
\end{equation}
this gas collapses and a liquid  forms.

At twice lower temperature
\begin{equation}
T_\lambda=\frac{2}{3k}\mathcal{E}_{x+y}%\label{f1}
\end{equation}
all zero-point oscillations become ordered. It creates an additional attraction and forms a single quantum ensemble.

A density of the boson condensate  is limited by  zero-point oscillations  of its atoms.
At condensation the distances between the atoms become approximately equal  to  amplitudes of zero-point oscillations.\\

Coming from it, we can calculate the basic properties of an ensemble of atoms with ordered zero-point oscillations, and compare them with measurement properties of superfluid helium.

We can assume that the radius of a helium atom is equal to the Bohr radius $a_B$, as it follows from quantum-mechanical calculations.
Therefore, the energy of electrons on the s-shell of this atom can be considered to be equal:
\begin{equation}
\hbar\Omega_0 = \frac{4e^2}{a_B}%\label{f1}
\end{equation}
As the polarizability of atom is approximately equal to its volume \cite{Fr}
\begin{equation}
A\simeq a_B^3,\label{vA}
\end{equation}
the potential energy of dispersive interaction (\ref{f2}), which causes the ordering zero-point oscillations in the ensemble of atoms, we can represent by the equation:
\begin{equation}
\mathcal{E}_{x+y} = - \frac{e^2}{a_B}a_B^6 n^2,\label{W}
\end{equation}
where the density of helium atoms
\begin{equation}
n=\frac{1}{L^3}
\end{equation}

\subsubsection{The velocity of zero-point oscillations of helium atom}
It is naturally  to suppose that zero-point oscillations of atoms are  harmonic  and the equality of kinetic and potential energies are characteristic for them:
\begin{equation}
\frac{M_4 \widehat{v_0}^2}{2} - \frac{e^2}{a_B}a_B^6 n^2 = 0,
\end{equation}
where $M_4$ is mass of helium atom, $\widehat{v_0}$ is their averaged velocity of harmonic zero-point oscillations.\\

Hence, after simple transformations we obtain:
\begin{equation}
\widehat{v_0}=c\alpha^3\left\{\frac{n}{n_0}\right\},\label{Ea31}
\end{equation}
where the notation is introduced:
\begin{equation}
n_0=\frac{\alpha^2}{a_B^3}\sqrt{\frac{M_4}{2m_e}}.\label{}
\end{equation}
If the expression in the curly brackets
\begin{equation}
\frac{n}{n_0}=1,\label{edin}
\end{equation}
we obtain
\begin{equation}
\widehat{v_0}=c\alpha^3\cong 116.5~
m/s.\label{alpha3}
\end{equation}

\subsubsection{The density of liquid helium}
The condition (\ref {edin}) can be considered as the definition of the density of helium atoms in the superfluid state:
\begin{equation}
n=n_0=\frac{\alpha^2}{a_B^3}\sqrt{\frac{M_4}{2m_e}}\cong 2.172\cdot 10^{22}~ atom/cm^3.\label{nnn}
\end{equation}
According to it, the density of liquid helium-4
\begin{equation}
\gamma_4 = n M_4\cong 0.144~ g/cm^3%\label{}
\end{equation}
that is in good agreement with the measured density of the liquid helium $0.145~g/cm^3$ for $T\simeq T_\lambda$.

Similar calculations for liquid helium-3 gives the density  $0.094~g/cm^3$, which can be regarded as consistent with its density $0.082 g/cm^3$ experimentally measured near the boiling point.

\subsubsection{The dielectric constant of liquid helium}
To estimate the dielectric constant of helium we can use
the Clausius-Mossotti equation \cite{Fr}:
\begin{equation}
\frac{\varepsilon-1}{\varepsilon+2}=\frac{4\pi}{3}{n}{A}.\label{KM}
\end{equation}
At taking into account Eq.(\ref{vA}), we obtain
\begin{equation}
\varepsilon\approx 1.040,
\end{equation}
that differs slightly from the dielectric constant of the liquid helium,
measured near the $\lambda$-point \cite{Russ}:
\begin{equation}
\varepsilon\approx 1.057
\end{equation}

\subsubsection{The temperature of $\lambda$-point}
The superfluidity is destroyed at the temperature $T_\lambda$, at which the energy of thermal motion is compared with the energy of the Van-der-Waals bond
in superfluid condensate:
\begin{equation}
\frac{3}{2}kT_\lambda - \frac{e^2}{a_B}a_B^6 n^2 = 0.
\end{equation}
With taking into account Eq.(\ref{nnn})
\begin{equation}
T_{\lambda}=\frac{1}{3k}\frac{M_4}{m_e}\frac{\alpha^4 e^2}{a_B}\label{f021}
\end{equation}
or after appropriate substitutions
\begin{equation}
T_{\lambda}=\frac{M_4 c^2 \alpha^6}{3k} =2.177 K,\label{f022}
\end{equation}
that is in very good agreement with the measured value $T_\lambda = 2.172K$.\footnote{There is a unexpected fact. The expression (\ref{f022}) for the temperature of $\lambda$-transition is given without any explanations in some articles of Internet at citing of patents \cite{Il}.  These articles and patents say nothing at all about zero-point oscillations, and  don't give generally any explanations of the reasons that allowed to write this expression.}

\subsubsection{The boiling temperature of liquid helium}
After comparison of Eq.(\ref{f1}) - Eq.(\ref{f2}), we have
\begin{equation}
T_{boil}=2T_\lambda=4.35 K\label{Tboil}
\end{equation}
This is the basis for the assumption that the liquefaction of helium is due to the attractive forces between the atoms with ordered
lengthwise  components of their oscillations.

\subsubsection{The velocity of the first sound in liquid helium}
It  is known from the theory of a harmonic oscillator that the maximum value of its velocity  is twice bigger than its average velocity.
In this connection, at assumption that the first sound speed  $c_{s1}$ is limited by this maximum speed oscillator, we obtain
\begin{equation}
c_{s1}=2\widehat{v_0}\simeq 233~m/s.\label{v01}
\end{equation}
It is in consistent with the measured value of the velocity of the first sound in helium, which has the maximum value of $238.3~m/s$ at $T\rightarrow 0$  and decreases with increasing temperature up to about $220~m/s$ at $T=T_\lambda$.\\

%\newpage
The results obtained in this section  are summarized  for clarity in the Table.(\ref{DT4}).\\
The measurement data in this table are mainly quoted by \cite{Russ} and \cite{Kik}.%\newpage
\begin{table}
\centering
\begin{tabular}{||c|c|c||c||}\hline\hline
&&&\\%\hline
  &defining &calculated&measured\\
  parameter&&&\\
  &formula&value&value\\
  &&&\\\hline
  the velocity of zero-point&&&\\
  oscillations of&$\widehat{v_0}=c\alpha^3$&$116.5$~&\\
  helium atom &&m/s&\\\hline\hline
  The density of atoms&&&\\
  in liquid &$n=\sqrt{\frac{M_4}{2m_e}}\frac{\alpha^2}{a_B^3}$&$2.172\cdot 10^{22}$&\\
  helium &&$atom/cm^3$&\\\hline\hline
  The density&&&\\
  of liquid helium-4&$\gamma=M_4 n$&$144.3$&$145_{T\simeq T_{\lambda}}$\\
   $ g/l$&&&\\\hline\hline
   The dielectric&&&$1.048_{T\simeq 4.2}$\\
  constant&$\frac{\varepsilon-1}{\varepsilon+2}=\frac{4\pi}{3}{\alpha^2}{\sqrt{\frac{M_4}{2m_e}}}$&1.040&\\
  of liquid helium-4&&&$1.057_{T\simeq T_{\lambda}}$\\\hline\hline
  The temperature &&&\\
  &$T_\lambda\simeq\frac{M_4 c^2 \alpha^6}{3}$&$2.177$&$2.172$\\
  $\lambda$-point,K&&&\\\hline\hline
  The boiling &&&\\
  temperature&$T_{boil}\simeq 2T_\lambda$&$4.35$&$4.21$\\
  of helium-4,K&&&\\\hline\hline
  The first sound &&&\\
  velocity,&$c_{s1}=2\widehat{v_0}$&$233$&$238.3_{T\rightarrow 0}$\\
  $m/s$&&&\\\hline\hline
  \end{tabular}
  \caption{Comparison of the calculated values of liquid helium with the measurement data}\label{DT4}
  \end{table}

\newpage
\subsection{The estimation of characteristic properties of He-3}
The results of similar calculations for the helium-3 properties are summarized in the Tab.(\ref{DT3}).

\begin{table}
\centering
\begin{tabular}{||c|c|c||c||}\hline\hline
&&&\\%\hline
 &defining &calculated&measured\\
  parameter&&&\\
  &formula&value&value\\
  &&&\\\hline
  The velocity of zero-point&&&\\
  oscillations of&$\widehat{v_0}=c\alpha^3$&$116.5$&\\
  helium atom &&m/s&\\\hline\hline
  The density of atoms&&&\\
  in liquid &$n_3=\sqrt{\frac{M_3}{2m_e}}\frac{\alpha^2}{a_B^3}$&$1.88\cdot 10^{22}$&\\
  helium-3 &&$atom/cm^3$&\\\hline\hline
  The density&&&\\
  of liquid&$\gamma=M_3 n_3$&$93.7~$&$82.3~$\\
  helium-3, g/l&&&\\\hline\hline
  The dielectric&&&\\
  constant&$\frac{\varepsilon-1}{\varepsilon+2}=\frac{4\pi}{3}{\alpha^2}{\sqrt{\frac{M_3}{2m_e}}}$&1.035&\\
  of liquid helium-3&&&\\\hline\hline
   The boiling&&&\\
   temperature&$T_{boil}\simeq \frac{4}{3}\frac{\mathcal{E}_{W}}{k}$&$3.27$&$3.19$\\
  of helium-3,K&&&\\\hline\hline
   The sound velocity&&&\\
  in liquid&$c_{s}=2\widehat{v_0}$&233~&\\
  helium-3&&m/s&\\\hline\hline
 \end{tabular}
 \caption{The characteristic properties of liquid helium-3}\label{DT3}
 \end{table}

\bigskip

There is a radical difference between mechanisms of transition to the superfluid state for He-3 and He-4.
Superfluidity occurs if complete ordering exists in the atomic system.
For superfluidity of He-3 electromagnetic interaction should order not only zero-point vibrations of atoms, but also the magnetic moments of the nuclei.

It is important to note that  all characteristic dimensions of this task: the amplitude of the zero-point oscillations, the atomic radius, the distance between  atoms in liquid helium - all equal to the Bohr radius $a_B$ by the order of magnitude.  Due to this fact, we can estimate the oscillating magnetic field, which a fluctuating electronic shell creates  on "its" \ nucleus:
\begin{equation}
H_\Omega\approx\frac{e}{a_B^2}\frac{a_B \Omega_0}{c}\approx \frac{\mu_B}{A_3},
\end{equation}
where $\mu_B=\frac{e\hbar}{2m_ec}$ is the Bohr magneton, $A_3$ is the electric polarizability of helium-3 atom.

Because the value of magnetic moments for the nuclei He-3 is approximately equal to the nuclear Bohr magneton $\mu_{n_B}=\frac{e\hbar}{2m_pc}$, the ordering in their system must occur below the critical temperature
\begin{equation}
T_c = \frac{\mu_{n_B}H_\Omega}{k}\approx  10^{-3} K.%.\label{f0}
\end{equation}
This finding is in agreement with the measurement data.
The fact that the nuclear moments can be arranged in parallel or antiparallel to each other is consistent with the presence of the respective phases of superfluid helium-3.\\

Concluding this approach permits to explain the mechanism of superfluidity in liquid helium.

In this way, the apposite  quantitative estimations of main parameters of the liquid helium and its transition to the superfluid state were  obtained.

 It was established that both related phenomena, superconductivity and  superfluidity, are based on the same physical mechanism: they both are consequences of the ordering of  zero-point oscillations.


\newpage
\begin{thebibliography}{33}\label{bibl}

\bibitem{BV1} Vasiliev B.V. : Physica C, {\bf{471}},277-284 (2011)

\bibitem{BV2} Vasiliev B.V. : Physica C, {\bf{483}},233-246 (2012)

\bibitem{BV3} Vasiliev B.V. : "Superconductivity, Superfluidity and Zero-Point Oscillations" \ in "Recent Advances in Superconductivity Research" \ , pp.249-280, Nova Publisher,NY(2013)

\bibitem{Landau} Landau L.D. : JETP, {\bf{11}}, 592 (1941)

\bibitem{Halat} Khalatnikov I.M.:  Introduction into theory of superfluidity , Moscow, Nauka, (1965)

\bibitem{FLondon}  London F.:  Trans. Faraday Soc. {\bf{33}}, p.8 (1937)

\bibitem{Fr} Fr\"{o}hlich H. : Theory of dielectrics, Oxford, 1957.


\bibitem{Russ} Russel J.Donnelly and Carlo F.Barenghy:
The Observed Properties of Liquid Helium, Journal of Physical and Chemical Data, $\bf{6}$, N1, pp.51-104, (1977)

\bibitem{Kik} Kikoine I.K. a.o.: Physical Tables, Moscow, Atomizdat (1978) (in Russian).


\bibitem{Il} Ilianok A.M: Eurasian patent № 003164, US patent 6,570,224B1, Korean patent N 10-0646267, China patent CN 1338120.
%\bibitem{MB} Born M.:  Mysterious number 137, Proc. Indian Acad. Sci. (A), 2, 533, (1935).


\end{thebibliography}
\end{document}